\documentclass[10pt,conference]{IEEEtran}
\IEEEoverridecommandlockouts
\usepackage{amsmath,amssymb,amsfonts}
\usepackage{algorithmic}
\usepackage[ruled]{algorithm2e}
\usepackage{array}
\usepackage{textcomp}
\usepackage{stfloats}
\usepackage{url}
\usepackage{color}
\usepackage{verbatim}
\usepackage{graphicx}
\usepackage{tabularx}
\usepackage{multirow}
\usepackage{array}
\usepackage{booktabs}
\usepackage{graphicx}
\usepackage{parskip}

\newtheorem{definition}{Definition}
\usepackage[numbers]{natbib}
\usepackage{bm}
\usepackage{mathtools}

\usepackage{subfigure}
\usepackage{booktabs}
\usepackage{multirow}
\usepackage{makecell}
\usepackage[table]{xcolor}

\usepackage{bibspacing}
\graphicspath{ {./fig/} }
\usepackage{enumitem}

\def\BibTeX{{\rm B\kern-.05em{\sc i\kern-.025em b}\kern-.08em
    T\kern-.1667em\lower.7ex\hbox{E}\kern-.125emX}}
\begin{document}

\title{Multi-View Trust Evaluation for Collaborator Selection via Evidential Deep Learning} 

\author{\IEEEauthorblockN{Botao~Zhu and Xianbin~Wang}
\IEEEauthorblockA{Department of Electrical and Computer Engineering, Western University,
London, Ontario, Canada}}

% \author{
% Botao~Zhu,~\IEEEmembership{Member,~IEEE} and Xianbin~Wang,~\IEEEmembership{Fellow,~IEEE
% }
% \thanks{
% This work was supported in part by the Discovery Program of Natural Sciences and Engineering Research Council of Canada (NSERC) under Grant
% RGPIN-2024-05720 and in part by the Canada Research Chair Program.

% B. Zhu and X. Wang are with the Department of Electrical and Computer Engineering, Western University, London, Canada N6A 5B9. Emails: \{bzhu88, xianbin.wang\}@uwo.ca.}
% }

% \markboth{}

\maketitle

\begin{abstract}

  % However, due to the inherent system complexity, observational trust-related data regarding collaborators is typically multi-source, heterogeneous, and of highly variable quality. 

Selection of trustworthy collaborators in distributed systems is critical for efficient task completion, necessitating the inference of trustworthiness from their past collaboration experience. However, as a collaborator serves distinct devices across diverse scenarios in past collaborations, its trust-related data, observed from different device-specific views, is inherently multi-source, heterogeneous, and uneven in quality. Consequently, achieving accurate trust evaluations for collaborator selection remains a major challenge. To tackle these issues, we propose a novel multi-view evidential learning (MVE) based trust evaluation method. First, to accommodate the multi-source heterogeneity of observed trust-related data, we model each task owner who has interacted with a potential collaborator as an independent observational view, enabling the evaluation of the collaborator's view-specific trust. Second, to address the dynamic evolution of trust under changing conditions, we leverage the powerful long-sequence modeling capability of the Mamba model to capture the deep temporal patterns of a collaborator's trust state within each view. Furthermore, to quantify the certainty levels of view-specific trust assessments, we incorporate an evidential deep learning mechanism in MVE, which outputs trust evaluation results while quantifying the subjective uncertainty underlying them. Finally, we employ a dynamic evidential fusion strategy to adaptively integrate the multi-view evidence based on their respective quantified uncertainties, thereby yielding a final trust evaluation for the collaborator. Extensive experiments demonstrate that the proposed MVE method outperforms baselines in both trust evaluation accuracy and task success rate. 
\end{abstract}


\section{Introduction}

% With the deep integration of communication, computation, and artificial intelligence, future systems are evolving toward highly collaborative intelligent platforms. In this paradigm, tasks 变得越来约复杂，例如（举例任务）， they are no longer executed by individual devices in isolation, but instead require dynamic orchestration of distributed resources and collaborative execution across devices. Accordingly, 选择可靠的合作者是重要的. However, as task的复杂性，系统的动态性，设备的异构型, traditional security- or privacy-based mechanisms insufficient for collaborator selection in future systems.

With the profound integration of communication, computation, and artificial intelligence, future networked systems are evolving toward highly collaborative platforms for intelligent resource utilization and goal realization. In this paradigm, tasks, such as cooperative perception and collective maneuver planning, have become increasingly complex, moving beyond the capabilities of isolated devices and requiring the dynamic orchestration of distributed resources and cross-device collaboration~\cite{Najib2024QSTrust}. Consequently, the fulfillment of task objectives critically depends on the selection of reliable collaborators. However, accurately selecting such collaborators remains challenging under the inherent task complexity, system dynamism, and device heterogeneity, rendering traditional security- or privacy-based mechanisms insufficient.

% In this context, trust has been recognized as a holistic tool for collaborator selection in collaborative systems. Specifically, trust is defined as the expectation of a task owner regarding a potential collaborator's capability and resources to successfully fulfill a given task~\cite{11296817}, evaluated from 信任相关的数据 of collaborators 在历史合作中-such as task success rate, latency, and 丢包率等. 然而，由于一个合作者信任相关的数据是在历史的交互中与不同的任务所有者合作在不同任务场景中所被采集的，这些数据天然就是（列举一些问题）由于任务的异构型和系统的动态性。不同的任务所有者对同一个合作者感知到的可信度根据它的信任相关的性能也是不一样的。所以这些数据是不能被合并到一起处理。

In this context, trust has been widely recognized as a holistic tool for collaborator selection in collaborative systems. Specifically, trust is defined as a task owner’s expectation of a potential collaborator’s capability and resources to successfully accomplish a given task~\cite{11296817}, typically evaluated from trust-related performance data, such as task success rate, latency, and packet loss. Such data is collected from a collaborator’s historical collaborations with multiple task owners across diverse tasks and dynamic environments. Since collaborations are recorded by different task owners under task-specific conditions, the data collected from these task owners constitutes distinct views of collaborator behavior, resulting in inherently multi-source and view-dependent trust-related data. Moreover, the observed trust-related data for the same collaborator may be uneven in quality and reliability across task owners, as it is influenced by varying task conditions and observation noise. Consequently, given heterogeneous trust-related data sources and imbalanced data quality, accurate assessment of collaborator trustworthiness in distributed systems requires solving the following challenges.

\textit{How to accurately evaluate trust from multi-source trust-related data?} In distributed systems, a collaborator’s performance is not a single static attribute, but a set of context-dependent manifestations shaped by task requirements, task owners, and fluctuating system conditions. For instance, a collaborator may perform reliably when interacting with one task owner under favorable conditions, yet exhibit degraded performance with another due to differences in network quality or task demands~\cite{11395598}. Consequently, its historical performance may appear substantially different from the views of these two task owners. If all historical records of a collaborator are aggregated for trust evaluation, the source structure of the historical records is discarded, obscuring contextual dependencies induced by task characteristics, collaboration relationships, and system states. This further prevents recognizing inconsistencies in observations across different task owners. Therefore, a more effective approach is to model task owners who have interacted with the same collaborator as multiple views, where each view represents a task-owner-specific trust evaluation of the collaborator, thereby distinguishing inter-view inconsistencies.

\textit{How to effectively fuse multi-view trust?} Although multi-view methods can distinguish trust assessments of a collaborator from different task owners, effectively fusing these views to derive a final trust decision remains challenging. First, the certainty of trust assessments for the same collaborator varies significantly across task owners. For example, task owners with frequent collaborations typically form more certain trust evaluations based on sufficient historical records, while those with sparse collaborations can provide only highly uncertain evaluations. As a result, a single scalar trust value is insufficient to characterize such assessments, as it fails to capture the underlying certainty. Second, in the view fusion stage, existing methods typically perform averaging or weighted fusion of different views, implicitly assuming that the trust evaluations of all views have equal certainty~\cite{han2023trusted}. However, when there is an imbalance in the degree of certainty among different views, this fusion mechanism cannot reflect the differences in certainty between views. Therefore, effective multi-view trust fusion should explicitly account for the certainty associated with each view and accordingly adjust their contributions in the final decision.

% To tackle the aforementioned challenges, we propose a novel multi-view evidential learning (MVE) based trust evaluation method for accurate and reliable collaborator selection. By assessing collaborators through diverse data sources across multiple views, our approach explicitly quantifies both the evaluated trust levels and their underlying uncertainties. These view-specific assessments are then dynamically fused to yield a final trust evaluation of collaborators. 

To tackle the aforementioned challenges, we propose a novel multi-view evidential learning (MVE)-based trust evaluation method for accurate and reliable collaborator selection. By evaluating collaborators from multiple views, MVE explicitly quantifies both trust evaluations and their associated uncertainties. These view-specific assessments are then dynamically fused to produce a final trust evaluation of each collaborator. The main contributions are summarized as follows.
\begin{itemize}[leftmargin=*]

\item We are the first to use multi-view framework for trust evaluation, enabling more effective utilization of diverse-source and varying-quality trust-related data in distributed systems.

    % 我们首次从多视角来用建模信任评估问题从而更好的使用来源多样和质量不一致的信任相关数据在分布式的系统中
    % \item We formulate trust evaluation as a multi-view  problem for the first time 解决了信任相关数据多源
    
    % fundamentally resolve the data heterogeneity and conflict challenges inherent in complex distributed systems.

    \item We propose the MVE method to implement the multi-view paradigm, which leverages Mamba to capture the distinct temporal dynamics within each view and resolves inter-view conflicts via a dynamic fusion mechanism.

    \item  We introduce evidential deep learning to shift trust evaluation from deterministic scoring to uncertainty-aware modeling. This explicitly quantifies evaluation confidence, empowering the system to filter unreliable evaluations and maintain superior robustness under conflict scenarios.
\end{itemize}

\vspace{-0.07 in}
\section{System Model and Problem Formulation}
\label{sec:system_model}

\subsection{System Model}

% We consider a distributed computing system composed of a set of $I$ heterogeneous devices $\mathcal{A} = \{a_1, a_2, \ldots, a_I\}$ together with a centralized trust server. Each device $a_i \in \mathcal{A}$ plays a dual role: it may act as a task owner that generates computation tasks requiring external assistance, or as a collaborator that accepts delegated tasks from other devices and executes them on their behalf. The trust server does not directly participate in task execution; instead, it is responsible for two core functions: (i) collecting and maintaining historical performance records of all collaborators, and (ii) providing trust evaluation and collaborator selection services for task owners upon the arrival of new tasks.

We consider a distributed computing system comprising a centralized trust server and a set of heterogeneous devices $\mathcal{A} = \{a_1, a_2, \ldots, a_I\}$. Each device operates with dual capabilities: serving as a task owner generating tasks, or as a collaborator processing delegated tasks. The trust server is dedicated to two core operations: (i) maintaining the historical performance records of all collaborators, and (ii) dynamically evaluating and selecting collaborators for task owners as new tasks arrive.

\textit{1) Task model}: We consider that device $a_i$, acting as a task owner, generates a task $\tau$ that is formally parameterized as $\tau = \bigl(\tau^{\text{size}},\, \tau^{\text{CPU}},\tau^{\text{thr}}, \tau^{\text{wi}}, \tau^{\text{tru}}\bigr)$,
where $\tau^{\text{size}}$ denotes the task size (MB), while $\tau^{\text{CPU}}$, $\tau^{\text{thr}}$, $\tau^{\text{wi}}$, and $\tau^{\text{tru}}$ specify the minimum requirements imposed on potential collaborators in terms of CPU frequency (GHz), throughput (Mbps), collaboration willingness, and trust level, respectively. It is worth noting that the task requirements can be extended as needed.

\textit{2) Task-specific trust model}: 
To identify a trustworthy collaborator, the task owner $a_i$ submits a task request to the trust server. The server then evaluates the trust of potential collaborators specific to the task $\tau$ by jointly considering their currently available resources and their historical performance records. Subsequently, the server recommends an optimal collaborator to the task owner $a_i$. Formally, the task-specific trust of potential collaborators is defined as follows.

\begin{definition}[Task-specific trust]
\label{def:trust}
Given the task $\tau$, the trust of a potential collaborator $a_j$, denoted as $T_{a_i \rightarrow a_j}(\tau) \in [0,1]$, is defined as the task owner $a_i$'s expectation of collaborator $a_j$'s ability and resources to successfully complete the task $\tau$, which is given by
% \vspace{-0.08 in}
\begin{equation}
\label{eq:trust_decomposition}
    T_{a_i \rightarrow a_j}(\tau) \;=\; T^{\text{hist}}_{a_j} \, T^{\text{res}}_{a_j}(\tau),
\end{equation}
where $T^{\text{res}}_{a_j}(\tau) \in [0,1]$ measures how well collaborator $a_j$'s currently available resources match the specific requirements of the task $\tau$, and $T^{\text{hist}}_{a_j} \in [0,1]$ characterizes the historical trustworthiness of collaborator $a_j$. If collaborator $a_j$ has previously collaborated with a set of task owners $\mathcal{V}_{a_j}$, its historical trust can be quantified based on its performance in these historical collaborations. 

% $T^{\text{hist}}_{a_j}$ is inferred from collaborator $a_j$'s historical performance records, accumulated through past collaborations with a set of task owners  is the set of task owners who have previously collaborated with collaborator $a_j$.

\end{definition} 

Based on the above model, the selected trustworthy collaborator undertakes the execution of task $\tau$. The detailed computation model is omitted due to space limitations and is referred to \cite{11296817}. 

\subsection{Problem Formulation}
\label{subsec:problem}
Given the task $\tau$, the objective of the system is to select the most trustworthy collaborator. The selection process is formulated as the following constrained optimization:
\vspace{-0.08 in}
\begin{alignat}{1}
     \label{problem}
     a_{j^\star} \;=\;& \arg\max_{a_j \,\in\, \mathcal{A}, \; a_j \neq a_i} \; T_{a_i \rightarrow a_j}(\tau) \\
    \mathrm{s.t.} \quad
    & T_{a_i \rightarrow a_j}(\tau) \,\geq\, \tau^{\text{tru}}, \\
    &|\mathcal{V}_{a_j}| \geq 1.
    % & T^{\text{hist}}_{a_j} \; \sim \; \bigcup_{a_m \in \mathcal{V}_{a_j}} \mathbf{R}_{a_j,a_m}.
\end{alignat}
Constraint (4) enforces that the selected collaborator must meet the minimum trust level required by the task $\tau$. Constraint (5) specifies that each potential collaborator has collaborated with at least one task owner, providing a minimum amount of data for trust evaluation.

% each potential collaborator $a_j$'s historical trust must be computed based on all its historical performance records, where $\mathbf{R}_{a_j,a_m}$ denotes its historical performance records when it collaborates with $a_m$, and 

\vspace{-0.05 in}
\section{MVE for Trust Evaluation}

\begin{figure*}[t!]
\centering
\includegraphics[scale=1]{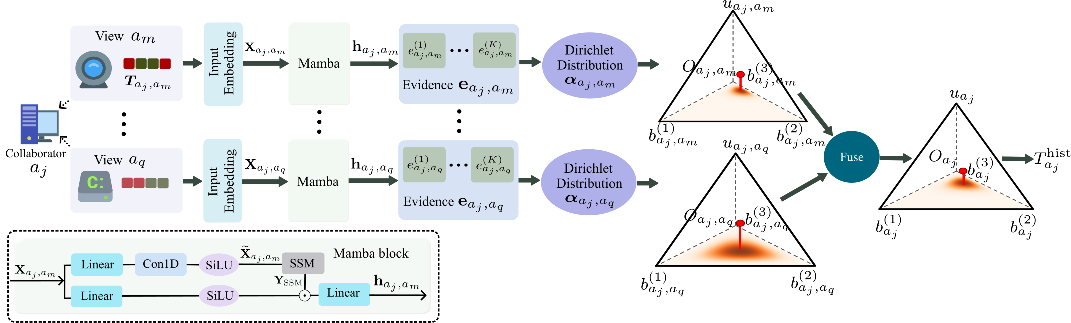}
\caption{An overview of the proposed MVE method for historical trust evaluation via multi-view temporal data structuring, uncertainty-aware evidential representation learning, and dynamic trust fusion.}
\label{workflow}
\end{figure*}

In this paper, we propose MVE to accurately evaluate the historical trust of potential collaborators from multiple views, addressing the challenges of multi-source and uneven-quality trust-related data, as shown in Fig.~\ref{workflow}. MVE first constructs a multi-view historical dataset for each collaborator, followed by temporal representation learning to capture view-specific trust dynamics. An uncertainty-aware evidential learning module is then employed to learn trust representations with quantified uncertainty. Finally, a multi-view evidential fusion mechanism aggregates all views to produce a robust historical trust score for each collaborator.

% \vspace{-0.1 in}
\subsection{Multi-View Historical Performance Dataset Construction}

Since the trust server monitors all collaborations, it maintains comprehensive historical records for all devices. 
For each collaborator $a_j$, let $\mathcal{V}_{a_j}$ denote the set of task owners that have previously collaborated with $a_j$. Because $a_j$ interacts with different task owners under different tasks and execution contexts, the historical records corresponding to each task owner capture a distinct observation view of $a_j$'s behavior. Therefore, each task owner $a_m \in \mathcal{V}_{a_j}$ is treated as an independent view, and the historical records from all such views are organized to construct a multi-view performance dataset for collaborator $a_j$. This formulation preserves the unique characteristics of each view, allowing the model to retain view-specific patterns and contextual information. Within view $a_m$, the historical performance records of collaborator $a_j$ are organized chronologically as a time-series sequence
$\mathbf{R}_{a_j,a_m} = \{ \mathbf{r}_{a_j,a_m}^{(n)}\}_{n=1}^{N_{a_j,a_m}}$,
where $N_{a_j,a_m}$ denotes the total number of collaborations between $a_j$ and $a_m$. Each record $\mathbf{r}_{a_j,a_m}^{(n)} = (s_{\text{comm}}^{(n)}, s_{\text{comp}}^{(n)})$ represents the communication and computation outcomes of the $n$-th collaboration, where $s_{\text{comm}}^{(n)}, s_{\text{comp}}^{(n)} \in \{0,1\}$, with $1$ indicating success and $0$ indicating failure. The trust score of collaborator $a_j$ in the $n$-th collaboration is computed as $T_{a_j,a_m}^{(n)} = s_{\text{comm}}^{(n)} \, s_{\text{comp}}^{(n)}$. Accordingly, the view-specific trust sequence is obtained as $\bm{T}_{a_j,a_m} = \{T_{a_j,a_m}^{(n)}\}_{n=1}^{N_{a_j,a_m}}$.

\vspace{-0.04 in}
\subsection{Temporal Trust Representation Learning via Mamba}
\label{subsec:temporal_encoding}
To obtain a view-level embedding of $\bm{T}_{a_j,a_m}$, Mamba is employed as the encoder. Compared with models such as LSTM and Transformer, Mamba achieves linear complexity with respect to sequence length and supports input-dependent state transitions via the selective state space mechanism (SSM), making it well-suited for modeling trust dynamics.

\textbf{Input embedding}:
Since $\bm{T}_{a_j,a_m}$ is a scalar sequence, each element is first projected into a $d$-dimensional embedding space via a learnable linear mapping:
\begin{equation}
\label{eq:input_projection}
    \mathbf{x}_{a_j,a_m}^{(n)} \;=\; \mathbf{W}_{\text{in}}\, T_{a_j,a_m}^{(n)} + \mathbf{b}_{\text{in}} \;\in\; \mathbb{R}^{d},
\end{equation}
where $\mathbf{W}_{\text{in}} \in \mathbb{R}^{d}$ and $\mathbf{b}_{\text{in}} \in \mathbb{R}^{d}$ are trainable parameters shared across all views. The embedded sequence is denoted by $\mathbf{X}_{a_j,a_m} = [\mathbf{x}_{a_j,a_m}^{(1)}, \ldots, \mathbf{x}_{a_j,a_m}^{(N_{a_j,a_m})}] \in \mathbb{R}^{d \times N_{a_j,a_m}}$. The embedded sequence $\mathbf{X}_{a_j,a_m}$ is then fed into a Mamba block. The processing pipeline of the Mamba block is summarized as follows.

\textbf{Input preprocessing}: The input $\mathbf{X}_{a_j,a_m}$ is first projected and processed by a causal convolution with SiLU activation:
% \vspace{-0.02 in}
\begin{equation}
    \widetilde{\mathbf{X}}_{a_j,a_m} \;=\; \operatorname{SiLU}\bigl(\operatorname{Conv1D}(\operatorname{Linear}(\mathbf{X}_{a_j,a_m}))\bigr).
\end{equation}
\textbf{Discretization}: The input-dependent SSM parameters are generated from $\widetilde{\mathbf{X}}_{a_j,a_m}$ and then are used to obtain the discretized state-transition matrices $\overline{\mathbf{A}}$ and $\overline{\mathbf{B}}$:
\vspace{-0.02 in}
\begin{align}
    \mathbf{B} &= \operatorname{Linear}(\widetilde{\mathbf{X}}_{a_j,a_m}), \quad \mathbf{C} = \operatorname{Linear}(\widetilde{\mathbf{X}}_{a_j,a_m}), \\
    \bm{\Delta} &= \operatorname{softplus}\bigl(\operatorname{Linear}(\widetilde{\mathbf{X}}_{a_j,a_m})\bigr), \\
    \overline{\mathbf{A}} &= \operatorname{Discrete}(\bm{\Delta}, \mathbf{A}), \quad \overline{\mathbf{B}} = \operatorname{Discrete}(\bm{\Delta}, \mathbf{A}, \mathbf{B}),
\end{align}
where $\operatorname{softplus}(\cdot)$ is a smooth approximation of ReLU that ensures the step size $\bm{\Delta}$ remains positive, and $\operatorname{Discrete}(\cdot)$ denotes the standard zero-order hold discretization.

\textbf{Selective scan and gated output}: The core SSM module $\operatorname{SSM}_{(\overline{\mathbf{A}}, \overline{\mathbf{B}}, \mathbf{C})}$ performs a selective scan over $\widetilde{\mathbf{X}}_{a_j,a_m}$ to produce the temporal output, which is then multiplicatively gated by a parallel branch and projected back to the original feature dimension:
\vspace{-0 in}
\begin{align}
    \mathbf{Y}_{\text{SSM}} &= \operatorname{SSM}_{(\overline{\mathbf{A}}, \overline{\mathbf{B}}, \mathbf{C})}\!\bigl(\widetilde{\mathbf{X}}_{a_j,a_m}\bigr), \\
    \mathbf{Y}_{a_j,a_m} &= \operatorname{Linear}\Bigl(\mathbf{Y}_{\text{SSM}} \odot \operatorname{SiLU}\!\bigl(\operatorname{Linear}(\mathbf{X}_{a_j,a_m})\bigr)\Bigr).
\end{align}
By adapting the step size $\bm{\Delta}$, the Mamba block selectively retains or discards historical information, enabling effective modeling of non-stationary trust dynamics. By stacking $L$ Mamba blocks, the output at the final step is taken as the view-level embedding $\mathbf{h}_{a_j,a_m} \in \mathbb{R}^{d}$, which captures the trust characteristics of collaborator $a_j$ from view $a_m$.

\vspace{-0.04 in}
\subsection{View-Specific Evidential Trust Learning}
\label{subsec:view_evidence}
Although the view-level embedding $\mathbf{h}_{a_j,a_m}$ produced by the Mamba encoder captures the trust dynamics of collaborator $a_j$ from view $a_m$, directly mapping it to a trust level via an activation function (e.g., softmax) would yield only a point estimate, leading to over-confident outputs and failing to express how confident view $a_m$ is about this evaluation. To address this limitation, we adopt the evidential deep learning~\cite{han2023trusted}. Specifically, for each view $a_m$, an opinion over collaborator $a_j$ is derived, jointly encoding both the evaluated trust level and the associated degree of certainty. These view-specific opinions are then fused to derive the final trust evaluation. In this subsection, we first present the view-specific trust learning process.

% \vspace{-0.05 in}
\textbf{Evidence generation}:
The embedding $\mathbf{h}_{a_j,a_m}$ from view $a_m$ is passed through a linear layer followed by a non-negative activation function to generate a $K$-dimensional evidence vector:
% \vspace{-0.1 in}
\begin{equation}
\label{eq:evidence_head}
    \mathbf{e}_{a_j,a_m} \;=\; \operatorname{Softplus} \bigl(\mathbf{W}_{\text{e}}\, \mathbf{h}_{a_j,a_m} + \mathbf{b}_{\text{e}}\bigr) \;\in\; \mathbb{R}_{\geq 0}^{K},
\end{equation}
where $\mathbf{W}_{\text{e}} \in \mathbb{R}^{K \times d}$ and $\mathbf{b}_{\text{e}} \in \mathbb{R}^{K}$ are the parameters, $K$ represents the number of trust levels, and $\operatorname{Softplus}(\cdot)$ guarantees that every component of $\mathbf{e}_{a_j,a_m}$ is non-negative. Each entry $e_{a_j,a_m}^{(k)} \in 
\mathbf{e}_{a_j,a_m}$ can be interpreted as the amount of evidence that view $a_m$ has collected for assigning $a_j$ to the $k$-th trust level.

% \vspace{-0.05 in}
\textbf{Encoding evidence as Dirichlet distribution}:
Following subjective logic~\cite{josang2016subjective}, the collected evidence is used to parameterize a Dirichlet distribution over the probability simplex of the trust label space:
% \vspace{-0.2 in}
\begin{align}
\label{eq:dirichlet_param}
      \operatorname{Dir}(\mathbf{p} | \bm{\alpha}_{a_j,a_m}) &= \frac{1}{B(\bm{\alpha}_{a_j,a_m})}\prod_{k=1}^{K}p^{\alpha^{(k)}_{a_j,a_m}-1}_{k}, \\
    \bm{\alpha}_{a_j,a_m} &= \mathbf{e}_{a_j,a_m} + \mathbf{1},
\end{align}
where $\mathbf{p} = [p_1,\ldots, p_{K}]$ denotes the vector of probabilities for $K$ different trust classes,  $\bm{\alpha}_{a_j,a_m} = [\alpha^{(1)}_{a_j,a_m}, \dots, \alpha^{(K)}_{a_j,a_m}]$ represents the Dirichlet parameters, $B(\boldsymbol{\alpha}_{a_j,a_m})$ is the multinomial Beta function that serves as a normalization constant, and $\mathbf{1}$ is the $K$-dimensional all-ones vector. The addition of $\mathbf{1}$ encodes a uniform base-rate prior that prevents the Dirichlet parameters from collapsing to zero in the absence of any evidence. The resulting distribution $\operatorname{Dir}(\mathbf{p}|\bm{\alpha}_{a_j,a_m})$ captures uncertainty in the trust evaluation by placing a distribution over the probability vector $\mathbf{p}$ parameterized by the observed evidence.

% \vspace{-0.05 in}
\textbf{Opinion modeling with belief and uncertainty}:
From the Dirichlet parameters, the subjective opinion $O_{a_j,a_m}$ of view $a_m$, consisting of a belief mass vector $\mathbf{b}_{a_j,a_m}$ and an uncertainty mass $u_{a_j,a_m}$, is derived as
% \vspace{-0.06 in}
\begin{equation}
\label{eq:belief_uncertainty}
    b_{a_j,a_m}^{(k)} \;=\; \frac{e_{a_j,a_m}^{(k)}}{S_{a_j,a_m}}, \qquad u_{a_j,a_m} \;=\; \frac{K}{S_{a_j,a_m}},
\end{equation}
where $S_{a_j,a_m} = \sum_{k=1}^{K} \alpha_{a_j,a_m}^{(k)} = \sum_{k=1}^{K} e_{a_j,a_m}^{(k)} + K$ is the Dirichlet strength. The belief mass $b_{a_j,a_m}^{(k)} \in [0,1]$ quantifies the support that view $a_m$ assigns to the $k$-th trust level, and the uncertainty mass $u_{a_j,a_m}\in (0,1]$ quantifies how much of the opinion remains unresolved due to insufficient evidence. Specifically, a larger $u_{a_j,a_m}$ indicates weaker evidence and lower confidence in the view-specific trust assessment, whereas a smaller $u_{a_j,a_m}$ indicates stronger evidence and a more confident opinion. It is straightforward to verify that $\sum_{k=1}^{K} b_{a_j,a_m}^{(k)} + u_{a_j,a_m} = 1$, so that the belief and uncertainty together form a valid mass assignment over the trust label space. The resulting opinion captures both the preference of view $a_m$ over trust levels and the associated uncertainty, providing a complete representation for subsequent multi-view fusion. Applying Eqs.~\eqref{eq:evidence_head}--\eqref{eq:belief_uncertainty} to each view $a_m \in \mathcal{V}_{a_j}$ yields a collection of view-specific opinions $\mathcal{O}_{a_j} = \{ O_{a_j,a_m} \}_{a_m \in \mathcal{V}_{a_j}}$. This set provides an uncertainty-aware characterization of collaborator $a_j$'s trustworthiness from multiple task-owner views.

% This set provides an uncertainty-aware characterization of how all task owners that have collaborated with collaborator $a_j$ perceive its trustworthiness.

% \vspace{-0.1 in}
\subsection{Evidential Multi-View Trust Fusion}
\label{subsec:fusion}
Given the view-specific opinion set $\mathcal{O}_{a_j}$, we proceed to aggregate these diverse opinions into a unified opinion that captures the collective trust assessment 
of collaborator $a_j$. We adopt the conflictive opinion aggregation 
rule proposed in~\cite{xu2024reliable}. For any two opinions $O_{a_j,a_m} = (\mathbf{b}_{a_j,a_m}, u_{a_j,a_m})$ 
and $O_{a_j,a_q} = (\mathbf{b}_{a_j,a_q}, u_{a_j,a_q})$, their fused opinion is calculated as 
% \vspace{-0.2 in}
\begin{align}
\label{eq:fused_belief_pair}
  O_{a_j} &= (\mathbf{b}_{a_j}, u_{a_j}) = O_{a_j,a_m} \oplus O_{a_j,a_q},
 \\
    b^{(k)}_{a_j} &= 
    \frac{b_{a_j,a_m}^{(k)} \, u_{a_j,a_q} \;+\; 
          b_{a_j,a_q}^{(k)} \, u_{a_j,a_m}}
         {u_{a_j,a_m} + u_{a_j,a_q}}, \\
          u_{a_j} &= 
    \frac{2\, u_{a_j,a_m}\, u_{a_j,a_q}}
         {u_{a_j,a_m} + u_{a_j,a_q}},
\end{align}
where $\oplus$ represents the fusion operation. As shown in Fig.~\ref{workflow}, the opinion $O_{a_j,a_m}$ from view $a_m$ exhibits a highly concentrated belief mass 
$\mathbf{b}_{a_j,a_m}=(0.05,0.15,0.70)$ with a low uncertainty $u_{a_j,a_m}=0.10$, indicating strong confidence in the high-trust level $b^{(3)}_{a_j,a_m} (0.70)$. In contrast, the opinion $O_{a_j,a_q}$ from view $a_q$ is characterized by a relatively flat belief mass 
$\mathbf{b}_{a_j,a_q}=(0.20,0.25,0.25)$ and a higher uncertainty $u_{a_j,a_q}=0.30$, reflecting the lack of a clear preference among the three trust levels. Applying the conflictive aggregation rule yields the fused opinion ${O}_{a_j}=({\mathbf{b}}_{a_j},{u}_{a_j})$, where ${\mathbf{b}}_{a_j}=(0.088,0.175,0.587)$ and 
${u}_{a_j}=0.150$. Despite the presence of a less informative view, the fused belief remains dominated by the third level (${b}^{(3)}_{a_j}=0.587$), demonstrating that the aggregation effectively preserves confident evidence while mitigating uncertainty.

For collaborator $a_j$ observed by $|\mathcal{V}_{a_j}|$ 
views, this rule is applied iteratively across all views to obtain 
the final fused opinion $\hat{O}_{a_j} = (\hat{\mathbf{b}}_{a_j}, 
\hat{u}_{a_j})$. For the fused opinion, the corresponding Dirichlet parameters 
$\hat{\boldsymbol{\alpha}}_{a_j}$ are reconstructed from 
$\hat{O}_{a_j}$. The projected probability of each trust level $k$ 
is computed following subjective logic~\cite{xu2024reliable}:
\begin{equation}
\label{eq:projected_prob}
    \hat{p}_{a_j}^{(k)} \;=\; \hat{b}_{a_j}^{(k)} \;+\; 
    \frac{1}{K}\, \hat{u}_{a_j},
\end{equation}
where the residual uncertainty is redistributed across all trust 
levels according to the uniform base rate. The historical trust of collaborator
$a_j$ is finally obtained by mapping each ordinal trust level 
$k \in \{1, \ldots, K\}$ to a numerical score $(k-1)/(K-1) \in [0,1]$ 
and taking its expectation under $\{\hat{p}_{a_j}^{(k)}\}_{k=1}^{K}$:
% \vspace{-0.1 in}
\begin{equation}
\label{eq:hist_trust}
    T^{\text{hist}}_{a_j} = \sum_{k=1}^{K} 
    \frac{k-1}{K-1} \hat{p}_{a_j}^{(k)}.
\end{equation}

% \vspace{-0.13 in}
\subsection{Loss Function}
To train the proposed MVE model, the loss function is formulated as
% \vspace{-0.1 in}
\begin{equation}
\label{eq:total_loss}
    \mathcal{L} = \mathcal{L}_{\text{acc}}(\hat{\boldsymbol{\alpha}}_{a_j}) + \beta \sum_{a_m \in \mathcal{V}_{a_j}} \mathcal{L}_{\text{acc}}(\boldsymbol{\alpha}_{a_j, a_m}) + \gamma \mathcal{L}_{\text{con}},
\end{equation}
where $\mathcal{L}_{\text{acc}}(\hat{\boldsymbol{\alpha}}_{a_j})$ and $\mathcal{L}_{\text{acc}}(\boldsymbol{\alpha}_{a_j, a_m})$ denote the evidential classification losses for the fused and individual views, respectively. Specifically, these losses utilize an expected cross-entropy term alongside an annealed Kullback-Leibler divergence regularizer. This combination minimizes the evidence generated for incorrect labels and prevents the premature convergence of misclassified instances to a uniform distribution. Furthermore, $\mathcal{L}_{\text{con}}$ serves as the consistency loss, which minimizes conflicts to explicitly ensure the alignment of trust evaluation results across different opinions during training. Finally, $\beta$ and $\gamma$ act as trade-off hyperparameters. For the complete mathematical derivation of the evidence-based loss and the conflictive penalty, readers are referred to~\cite{xu2024reliable}.

\section{Resource Evaluation and Collaborator Selection}
After obtaining the historical trustworthiness of all potential collaborators, the task owner $a_i$ further evaluates the resource trustworthiness of each collaborator for the specific task $\tau_{a_i}$. Following the task requirements defined in Eq. (1), we collect the real-time resource status from each potential collaborator $a_j$, including its available storage size $a^{\text{size}}_j$, CPU frequency $a^{\text{CPU}}_j$, throughput $a^{\text{thr}}_j$, and collaborative willingness $a^{\text{wi}}_j$. By comparing each dimension of the collaborator's status with the corresponding task requirement, the resource trustworthiness $T^{\text{res}}_{a_j}$ is calculated using a multiplicative approach as follows:
\vspace{-0.05 in}
\begin{equation}
T^{\text{res}}_{a_j}(\tau) = \prod_{f \in \{\text{size, CPU, thr, wi}\}} \Phi_f,
\label{resource_evaluation}
\end{equation}
where $\Phi_f \in \{0, 1\}$ denotes the satisfaction degree for each dimension. Based on each collaborator $a_j$'s historical trust $T^{\text{his}}_{a_j}$ and resource trust $T^{\text{res}}_{a_j}$, the task owner $a_i$ selects the collaborator with the highest trust value $T_{a_i \to a_j}$ as the final collaborator.

% \vspace{-0.15 in}
\section{Result Analysis}

\subsection{Experimental Setup}
We consider two tasks with different computational intensities, namely face recognition and virus scanning, with the same default task size of 500 MB~\cite{11296817}. Devices DELL 5200, DELL 5820, and DELL 7060 are employed, where HeavyLoad and Clumsy are used to control their computational and network resources, respectively, thereby defining five representative operational scenarios: elite, stable, strategic, selfish, and failed. Following \cite{Favour2025Benchmarking}, each device executes 200 task instances under each scenario, and a scenario-specific device model is constructed from the extracted device behaviour features. The five scenarios are mapped to a five-level trust taxonomy serving as ground-truth annotations, where elite, stable, strategic, selfish, and failed correspond to highly trusted $[0.9, 1.0)$, trusted $[0.7, 0.9)$, non-stationary $[0.4, 0.7)$, constrained $[0.2, 0.4)$, and malicious $[0.0, 0.2)$, respectively. From the extracted behavioral features, a total of 15 distinct scenario-specific device models (3 devices $\times$ 5 scenarios) are constructed. A collaborative system consisting of 200 simulated devices and 1 server is constructed using Python and NS-3, where each device is instantiated based on the 15 scenario-specific device models. A total of 10,000 collaborative tasks are randomly executed to collect interaction traces, which are subsequently used for model training and performance evaluation. The Mamba module comprises $3$ Mamba blocks with a hidden dimension $d=64$. The proposed MVE is trained on the Lambda workstation, with a learning rate of $10^{-3}$, a weight decay of $10^{-4}$, and a dropout rate of 0.1. The hyperparameters $\beta$ and$\gamma$ in the loss function are set to 1.

\subsection{Macro-F1 and MAE Comparisons}
To validate the effectiveness of the proposed MVE, Macro-F1 and mean absolute error (MAE) are employed as performance metrics to benchmark all methods on the historical trust evaluation of collaborators. For fair comparison, the deep evidential fusion network (DEF) and trusted multi-view (TMC) are implemented within the same framework as MVE, with only the neural network backbone and the fusion mechanism replaced by their respective counterparts. As presented in Table~\ref{tab:overall_comparison}, QS-Trust achieves the lowest performance, and TMC enhances the results by integrating evidential learning with Dempster's rule for multi-view fusion. DEF further advances the results through its evidence discount mechanism that down-weights low-quality views before applying Dempster's rule. Ultimately, the proposed MVE achieves the best performance on both metrics, outperforming DEF by 4.86\% in Macro-F1 and 14.6\% in MAE. The empirical results consistently demonstrate the superiority of the proposed MVE in reliably evaluating historical trust of collaborators.

% \vspace{-0.05 in}
\subsection{Uncertainty Analysis and Robustness to Conflict}
\begin{figure}[!t]
      \centering
      \subfigure[]{\includegraphics[scale=0.58]{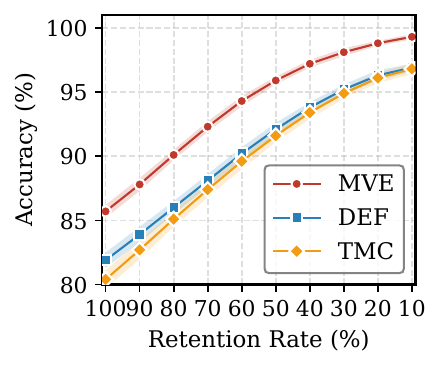}}
      \hspace{-0.01 in}\subfigure[]{\includegraphics[scale=0.6]{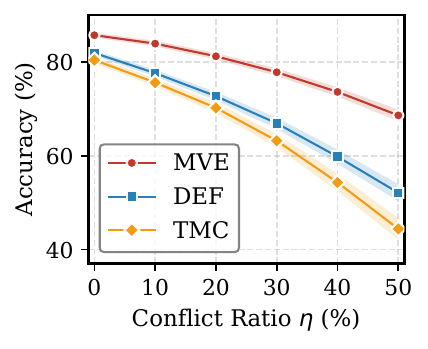}}
      \caption{(a) MVE provides the most discriminative uncertainty estimates to achieve higher accuracy. (b) Accuracy under varying inter-view conflict ratios, highlighting the superior robustness of the proposed MVE over baseline methods.}
     \label{cetentaity_fuse}
\end{figure}

 To verify whether the uncertainty is informative for identifying unreliable evaluations, we proceed as follows. For each test sample, we obtain its predicted trust level and the fused uncertainty mass $\hat{u}_{a_j}$. Test samples are sorted in ascending order of $\hat{u}_{a_j}$, and only the top-$r$ fraction with the lowest uncertainty is retained, where $r$ varies from 100\% to 10\%. The accuracy of collaborators' historical trust evaluation within the retained subset is recomputed at each retention rate. As shown in Fig.~\ref{cetentaity_fuse}~(a), all three methods exhibit monotonically increasing accuracy as high-uncertainty samples are progressively removed. MVE achieves the steepest ascent--from 85.7\% to 99.3\%--demonstrating the strongest correlation between its predicted uncertainty and actual prediction errors. DEF and TMC also show upward trends but plateau earlier at 96.9\% and 96.8\% respectively, indicating that their Dempster-based fusion produces less discriminative uncertainty estimates. This result has practical implications for collaborator selection: task owners can leverage MVE's uncertainty to filter out unreliable evaluations. 
 
 % \vspace{-0.1 in}
To examine the robustness of MVE under inter-view conflict, we simulate collusive misreporting scenarios on the test set. For a collaborator, a fraction $\eta$ of its views is randomly selected and all entries in the corresponding records are bit-wise flipped, so that these views convey opinions opposite to the true trust. The conflict ratio $\eta$ varies from 0\% to 50\%. Conflict is injected only at inference time, while the training phase uses clean data. As shown in Fig.~\ref{cetentaity_fuse}~(b), as $\eta$ increases, TMC exhibits the most severe degradation, which is consistent with the known limitation of Dempster’s rule, where high conflict mass results in unstable and unreliable fusion. DEF mitigates this to some extent through its evidence discount mechanism. In contrast, the proposed MVE degrades by only 17.1\%, maintaining an accuracy of 68.6\% at $\eta = 50\%$, thereby demonstrating superior robustness under high-conflict conditions.

\begin{table}[t]
\centering
\caption{Performance comparison of MVE and baselines on historical trust evaluation. Results are reported over 5 runs.}
\label{tab:overall_comparison}
\renewcommand{\arraystretch}{1.2}
\setlength{\tabcolsep}{6pt}
\begin{tabular}{l|cc}
\toprule
 \textbf{Method} & \textbf{Macro-F1} $\uparrow$ & \textbf{MAE} $\downarrow$ \\
\midrule
  QS-Trust~\cite{Najib2024QSTrust}                & $0.5934 \pm 0.0157$ & $0.1635 \pm 0.0079$ \\
  TMC~\cite{han2023trusted} & $0.7693 \pm 0.0072$ & $0.0912 \pm 0.0038$ \\
  DEF~\cite{xu2022deep}                & $0.7865 \pm 0.0068$ & $0.0859 \pm 0.0036$ \\
\rowcolor{red!15}
\textbf{MVE} & $\mathbf{0.8247 \pm 0.0058}$ & $\mathbf{0.0734 \pm 0.0029}$ \\
\bottomrule
\end{tabular}
\end{table}

% \vspace{-0.01 in}
\subsection{Task Success Rate Comparison Under Varying Requirements}

\begin{figure}[t!]
\centering
\includegraphics[scale=0.79]{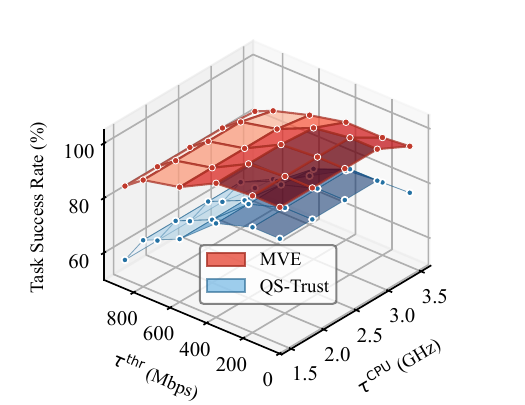}
\caption{The proposed MVE achieves higher task success rates than QS-Trust.}
\label{fig:tsr_3d}
\end{figure}

Fig.~\ref{fig:tsr_3d} presents the task success rate comparison when varying the task requirements $\tau^{\text{thr}}$ and $\tau^{\text{CPU}}$. As task requirements increase, MVE yields a significantly higher task success rate than QS-Trust, with the performance gap widening progressively. This superiority stems from the task-specific resource trust evaluation $T^{\text{res}}_{a_j}(\tau)$ defined in Eq.~(\ref{resource_evaluation}). Rather than relying on a coarse-grained, generic resource score, our approach performs a rigorous per-dimension binary check against specific task demands. Consequently, it precisely filters out candidate devices whose CPU frequency, bandwidth, or other critical capabilities fall short, thereby facilitating the selection of the most suitable collaborator.

% \vspace{-0.09 in}
\section{Conclusion}

In this paper, we propose the MVE learning framework to tackle the critical challenges of multi-source data and inconsistent observational quality in trust evaluation. By synergistically integrating multi-view learning with the uncertainty quantification mechanisms of evidential deep learning, MVE shifts the paradigm of trust assessment from vulnerable deterministic scoring to robust, evidence-based fusion. Extensive evaluations validated that MVE explicitly identifies and filters unreliable observations, yielding significant improvements in trust evaluation accuracy and task success rate.

% \vspace{0.12 in}
\footnotesize

\end{document}